\documentstyle{article}
\font\tenbf=cmbx10
\font\tenrm=cmr10
\font\tenit=cmti10
\font\elevenbf=cmbx10 scaled\magstep 1
\font\elevenrm=cmr10 scaled\magstep 1
\font\elevenit=cmti10 scaled\magstep 1

\newcommand{\appa}{\mbox{\ae}}

\def\lsim{\mathrel{\rlap{\lower3pt\hbox{\hskip0pt$\sim$}}
    \raise1pt\hbox{$<$}}}         
\def\gsim{\mathrel{\rlap{\lower4pt\hbox{\hskip1pt$\sim$}}
    \raise1pt\hbox{$>$}}}         

\newcommand{\eeq}{\end{equation}}
\newcommand{\beq}{\begin{equation}}

\renewcommand{\Im}{\mbox{Im}\,}
\renewenvironment{thebibliography}[1]
 { \elevenrm
   \begin{list}{\arabic{enumi}.}
    {\usecounter{enumi} \setlength{\parsep}{0pt}
     \setlength{\itemsep}{3pt} \settowidth{\labelwidth}{#1.}
     \sloppy
    }}{\end{list}}

\begin{document}
\begin{flushright}
\vspace*{-.15in}
\Large
UND-HEP-92-BIG\hspace*{.1em}08\\
\large
November 1992
\end{flushright}
\vspace*{1cm}

\begin{center}{{\tenbf {\em\bf FSI } PHASES AND {\em\bf CP } ASYMMETRIES IN
BEAUTY: \\
               \vglue 10pt
               {\em \bf QCD } POINT OF VIEW \footnote{\large {Talk given at
DPF-92 meeting, FERMILAB, November 10-14, 1992.}}\\}
\vglue 5pt
\vglue .8cm
{\tenrm N.G.URALTSEV \\}
\baselineskip=13pt
{\tenit Physics Department, University of Notre Dame\\}
\baselineskip=12pt
{\tenit Notre Dame, IN 46556, USA\\}
\vglue 0.15cm
{\tenrm and\\}
\vglue 0.15cm
{\tenit St.Petersburg Nuclear Physics Institute\\}
\baselineskip=12pt
{\tenit Gatchina, St.Petersburg District, 188350, Russia\\}
\vglue 0.9cm
{\tenrm ABSTRACT}}
\thispagestyle{empty}
\end{center}
\vglue 0.3cm
{\rightskip=3pc
 \leftskip=3pc
 \tenrm\baselineskip=12pt
 \noindent
Inclusive and exclusive decays of heavy flavours look quite different from
a theoretical viewpoint. We argue that inclusive decays can be treated
quantitatively and report on the calculation of the perturbative
corrections to the Final State Interaction phases
in connection to their impact
on effects of {\elevenit CP} violation.

\vglue 0.7cm}
\elevenrm

Present expectations for the observation of the {\elevenit CP}
nonconservation in
beauty are based on the search for those {\elevenit CP} odd
effects in $B$ mesons
that appear due to $B^0-\bar{B}^0$ mixing. Some modes -- like
$B_d\rightarrow
\psi\, K_s\:,\,\pi^+\pi^-$ have the additional advantage of being
``clean''
from a theoretical point of view. Of much practical interest is
another kind of effects where the $B^0 -\bar{B}^0$  mixing is not
important, say in decays of $B^{\pm}$   or $\Lambda_b$.   However
the simplest {\elevenit CP} odd asymmetries in such decays
are rather obscure as they depend crucially on the
{\elevenit CP} even phases generated by the strong
interactions in the final state (the FSI phases). (For a general review
see e.g. ref.$^1$).

The decays mediated by the $b\rightarrow u \bar{u} s$ transitions
look most promising.
The  main  reason  is that here
the  ``Penguin''   amplitudes   due   to  the $b
\rightarrow s+(c\bar{c},t\bar{t})_{virt} \rightarrow s+q \bar{q}$
chain also contribute,
and their magnitude must be close to the strength of the doubly KM suppressed
tree
level  amplitude$^2$. The relevant Penguin amplitude
has  literally  a {\elevenit CP} even phase
$\delta_P$, $\tan \delta_P \simeq  \pi /\log \frac{m_t^2,M_W^2}{m_b^2}$, or
numerically $\delta_P \simeq 0.5$;  however  including
the sizeable mass of the $c$ quarks reduces the estimate to
$\delta_P\simeq 0.1$ .

  There is a popular opinion that other FSI phases are negligible and one
can rely on the Penguin
phase $\delta_P$. Simultaneously
the  opposite point of view can be found in the literature -- that
in spite of the large $b$ quark mass and
energy release  the FSI phases are generally
large in heavy flavour decays. In such a case  however
it is hardly possible to predict
theoretically even
the sign of the effect.

  We believe$^3$ that {\elevenit a priori} there are no sound grounds  to
consi
der
FSI phases as small, at least  as  compared  to  $\delta_P$.  This
should  be especially the case for a generic exclusive process  where  the
result  depends crucially not only on the hadronization details but also on
the actual dynamics of  the  formation
of  the  particular final state at the ``hard'' stage$^2$.
For instance  for  color suppressed decays the hard part of the
process can naturally  involve a hard gluon exchange; if so one
might expect the FSI phase to be  of the order of $\pi /2$.
On the contrary for inclusive processes
the QCD language of quarks and  gluons is adequate$^4$:  perturbative
corrections  are  governed  by  the parameter
$\alpha_s(m_b^2)$ and nonperturbative effects  are power-like suppressed;
therefore here
the $\delta_P$ based  estimate is a reasonable
first approximation.

  Exploring this idea we concentrated$^3$ on
 the {\elevenit CP} odd inclusive width difference
for the decays of $b$ and $\bar{b}$ quarks into the states without  heavy
quarks.   We calculated higher order QCD effects and
found them to reduce the asymmetry only slightly (by 10-20\%). This
disagreed with the  result  of the paper$^5$ where the strong
cancellation of the effect had been claimed.

We start
with the main equation for the {\elevenit CP} odd width difference
\beq
\Delta\Gamma \equiv \Gamma (\bar{b}\rightarrow \bar{s} q \bar{q})
- \Gamma (b\rightarrow sq \bar{q}) = -4
\Im(\lambda_i\lambda_j^*)\cdot\sum_F \Im(A_i(b\rightarrow F)\cdot
A_j^*(b\rightarrow F)),
\eeq
where $F$ are the final states included
for the process, $A_{i,j}$   are the decay amplitudes with  the  KM
factors  factored  out  and $\lambda_{i,j}$  are  the corresponding
KM factors. The $S$ matrix unitarity enables  one  to  express the absorptive
parts of the amplitudes $A$
as a sum over the set $I$ of real
intermediate states for the $b\rightarrow F$  transitions:
\[
\Delta\Gamma=-2 \Im(\lambda_i\lambda_j^*)\cdot\sum_F \sum_I
\{A_i(b\rightarrow I) A^*(F\rightarrow I) Re
A_j(b\rightarrow F)-  \]
\beq    - A_j(b\rightarrow I)
A^*(F\rightarrow I) Re A_i(b\rightarrow F)\} ,
\eeq
with $A(F\rightarrow I)$ being generated by strong interactions.

In the  perturbative QCD expansion $\Delta\Gamma$  appears  in  order
$\alpha_s$ and does not contain $\log(m_t^2/m_b^2)$. In the
spirit of the standard  LLA  we  have calculated  all  the
corrections  of  the form $\alpha_s^{n+1}\,\log^n(m_t^2/m_b^2)$. The
result  appears  to  be  very  simple.  To calculate
$\Delta\Gamma$ one should consider only the one-gluon
rescattering amplitude  $c \bar{c} X \rightarrow q \bar{q} X'$,
thus almost reproducing the lowest order
estimate based on the Penguin phase $\delta_P$. The account for the
higher orders reduces merely to using the  weak  decay
amplitudes $A_{i,j}(b\rightarrow F,I)$ obtained by the {\elevenit effective}
$\Delta B=1$ weak  interaction Lagrangian normalized at the scale
$q^2=-m_b^2$ rather than by the bare one defined at
$q^2=-M_W^2$.

To prove this prescription one can consider all possible states $F$
and $I$ (such as $s+g$, $s q \bar{q}$, $s q \bar{q}+g$ \ldots) step by step
 and count the powers of $g_s$
in the the corresponding  `strong'  amplitudes $A(F\rightarrow
I)$. The important point here
is  that  owing  to  the on-shellness of  both $F$  and  $I$ the
strong amplitude
$A(F\rightarrow I)$  cannot  contain large logs, $\log
\frac{m_t^2,M_W^2}{m_b^2}$, provided it is expressed in
terms  of the $\alpha_s$ normalized at $q^2 =-m_b^2$.
In fact this statement is nothing but the renormalizability of QCD.

The states $I$ without a  $c\bar{c}$  pair cannot
contribute to $\Delta\Gamma$. The rescattering amplitude
$A(F\rightarrow I)$ therefore describes a transition from a charmless state
to a state with hidden charm and so it contains at least one
power of $\alpha_s$. For that reason the leading contribution is obtained
only if the single gluon amplitude is considered whereas in the remaining weak
amplitudes $A_{i,j}$ any extra power of the strong coupling is accompained by
the `ultraviolet' {\elevenit log}. The letter requirement defines exactly
the `standard' LLA renormalized weak decay amplitudes.

  Actually the renormalization of the weak amplitudes not only modifies their
strength but also makes new states possible on the quark level -- say
$\bar{d}d$ can now appear as well as
$\bar{u}u$; all they should be taken
into account. The following observation$^3$ simplifies the analysis further:
the states like $gg+s$, $ggg+s$ {\elevenit etc.} {\elevenit
are absent} in the LLA.

Indeed, the Penguin operator which appears in the effective
Lagrangian  due  to  the integration over  the  virtual  states
with $q^2\gg m_b^2$ has the form
$\bar{b}\gamma_{\mu}\frac{\lambda^a}{2}s\cdot
\nabla_{\nu}G_{\mu\nu}^a$.
In the absence of the light quark legs the equation of
motion $\nabla_{\nu}G_{\mu\nu}^a =0$ tells the vanishing of
its matrix element  for  purely gluon final states in the
leading order in $\alpha_s$. For the $gg+s$ states, in
particular,  it corresponds to the LLA cancellation for
real gluons of the contributions  of two possible graphs.

 Finally the  first  nontrivial  correction  to
$\Delta\Gamma$ generated in two loops reads as
\begin{eqnarray}
\Delta\Gamma& \propto &\frac{\alpha_s}{3\pi}\Im
[\log\frac{m_t^2}{m_b^2}
+i\pi\zeta(m_c^2/m_b^2)] \rightarrow \nonumber \\
 &\rightarrow &
\frac{\alpha_s(m_b^2)}{3\pi}\Im [\log\frac{m_t^2}{m_b^2}
 +i\pi\zeta(m_c^2/m_b^2)
\cdot(1-\frac{\alpha_s}{3\pi}(n_f+1)\log
\frac{m_t^2}{m_b^2}+\appa(c_{\pm}))]\;,
\end{eqnarray}
where $n_f$  is the number of light flavours excluding $c$
quark, $\zeta\simeq 0.2$ is the phase space suppression factor,
$\appa\simeq 2\alpha_s/4\pi \cdot \log M_W^2/m_b^2$  is the
`ordinary' correction factor due to the  renormalization of the
standard color factors $c_{\pm}$   in the effective Lagrangian. (In a sense the
last term
$\appa$ in the eq.(3)  together  with the unity added to $n_f$
can be attributed to the moduli of the
interfering amplitudes rather than to  the  phases).  For  $n_f=3$
and $\alpha_s(m_b^2)=0.18$ the first term (representing the less
trivial correction) is about $-0.17$, however it is strongly
canceled by the trivial corrections, $\appa\simeq 0.13$. The
summation of all orders in the LLA practically does not change the
conclusion: the total correction to $\Delta\Gamma$ if taken
literally appears to be near $-2\%$ for $\Lambda_{QCD}= 0.1\div 0.3
GeV$.  Numerically for the  inclusive  asymmetry  one has
\beq
\frac{\Gamma(\bar{b}\rightarrow \bar{s}+charmless)-\Gamma(b\rightarrow
 s+charmless)}
{\Gamma(\bar{b}\rightarrow \bar{s}+charmless)+\Gamma(b\rightarrow
 s+charmless)}
\simeq -1.9\cdot\zeta\cdot\left|\frac{V_{ub}}{V_{cb}}\right|\cdot\sin\alpha
 \simeq -2\cdot 10^{-2}\;,
\eeq
where $\alpha=arg(V_{cb}^*V_{cd}V_{ub}V_{ud}^*)$
is one of the  angles  of  the
Unitarity Triangle$^1$, $\alpha\simeq 0.55$  at
$|V_{ub}/V_{cb}|=0.1$ if one assumes $f_B\simeq 140\;MeV$.  The total
probability of such decays
$BR(b\rightarrow s+charmless)$  is  in this case about
$2.5\cdot10^{-3}$.

In the similar way one calculates also the {\elevenit CP}
odd  inclusive width differences in the decays
$b\rightarrow qs\bar{s}\,$, which  are  nonzero owing to the different
phase  space  for  the  $c\bar{c}$ and  $u\bar{u}$ intermediate states. Under
the same numerical assumptions the rate
asymmetry  for $b\rightarrow ss\bar{s}$  is  about
$-1.8\cdot 10^{-2}$ at $BR(b\rightarrow ss\bar{s})\simeq 5\cdot10^{-4}$
and for
the channel $b\rightarrow ds\bar{s}$ the asymmetry    is    near
$+1.3\cdot10^{-1}$ while $BR(b\rightarrow ds\bar{s})\simeq 7\cdot 10^{-5}$.

The measurement of the inclusive {\elevenit CP} odd asymmetries
in beauty particles seems to be an extremely difficult experimental
problem and similar analysis
for exclusive few body modes
would be more relevant for experiment. Nevertheless
the QCD calculations we have made are not useless.
For these corrections enter any quark diagram that
proceeds via Penguins. Our caution to apply directly similar calculations
to exclusive decays is in fact no more but the
statement that in that case there could be other and even larger
sources for the FSI phases.
\vglue 0.5cm
{\elevenbf \noindent Aknoledgements \hfil}
\vglue 0.5cm
I am grateful to my coauthor on paper$^3$ Yu.Dokshitzer for
collaboration and to I.Bigi for pursuing me to address the problem of FSI
phases from the positions of QCD. This work was supported in part by National
Science Foundation under grant number NSF-PHY 92 13313.
\vglue 0.5cm
{\elevenbf\noindent References \hfil}
\nopagebreak
\vglue 0.35cm

\end{document}